\documentclass[10pt, conference, letterpaper]{IEEEtran}
\IEEEoverridecommandlockouts
\usepackage{cite}
\usepackage{amsmath,amssymb,amsfonts}
\usepackage{amsthm}
\usepackage{algorithm}  
\usepackage{algorithmicx}  
\usepackage{algpseudocode}
\usepackage{graphicx}
\usepackage{epstopdf}
\usepackage{textcomp}
\usepackage{xcolor}
\usepackage[T1]{fontenc}
\usepackage{caption}
\usepackage{cases} 
\usepackage{amssymb}
\usepackage{tabulary}
\usepackage{url}
\usepackage{booktabs}
\usepackage{diagbox}
\usepackage{amsmath}
\usepackage{bm}
\usepackage{tablefootnote}
\usepackage{float} 
\usepackage{subcaption}
\usepackage{color}
\usepackage{color,xcolor}

\usepackage{subcaption}
\usepackage{url}

\usepackage{pifont}
\usepackage{xcolor}

\usepackage{listings}
\usepackage{xcolor}

\def\BibTeX{{\rm B\kern-.05em{\sc i\kern-.025em b}\kern-.08em
    T\kern-.1667em\lower.7ex\hbox{E}\kern-.125emX}}
\begin{document}
\lstset{
  language=Python,               
  basicstyle=\ttfamily\scriptsize, 
  keywordstyle=\color{blue},     
  commentstyle=\color{gray},     
  stringstyle=\color{red},       
  numbers=left,                  
  numberstyle=\tiny\color{gray}, 
  stepnumber=1,                  
  numbersep=5pt,                 
  backgroundcolor=\color{white}, 
  showspaces=false,              
  showstringspaces=false,       
  showtabs=false,                
  frame=none,                  
  tabsize=4,                    
  captionpos=b,                 
  breaklines=true,               
  breakatwhitespace=false,       
  escapeinside={\%*}{*)}         
}

\title{LLMSched: Uncertainty-Aware Workload Scheduling for Compound LLM Applications\\
}

\author{\IEEEauthorblockN{Botao Zhu\textsuperscript{1}, Chen Chen\textsuperscript{2}, Xiaoyi Fan\textsuperscript{3}, Yifei Zhu\textsuperscript{1}\textsuperscript{*}}
\IEEEauthorblockA{\textsuperscript{1}UM-SJTU Joint Institute, Shanghai Jiao Tong University, Shanghai, China \\
\textsuperscript{2}John Hopcroft Center for Computer Science, Shanghai Jiao Tong University, Shanghai, China \\
\textsuperscript{3}Jiangxing Intelligence Inc., Shenzhen, China \\
Email: \{zhubotao, chen-chen\}@sjtu.edu.cn, xiaoyi.fan@ieee.org, yifei.zhu@sjtu.edu.cn
\thanks{\textsuperscript{*}Corresponding author.}}
}

\maketitle

\begin{abstract}
Developing compound Large Language Model (LLM) applications is becoming an increasingly prevalent approach to solving real-world problems. In these applications, an LLM collaborates with various external modules, including APIs and even other LLMs, to realize complex intelligent services. However, we reveal that the intrinsic duration and structural uncertainty in compound LLM applications pose great challenges for LLM service providers in serving and scheduling them efficiently. In this paper, we propose LLMSched, an uncertainty-aware scheduling framework for emerging compound LLM applications. In LLMSched, we first design a novel DAG-based model to describe the uncertain compound LLM applications. Then, we adopt the Bayesian network to comprehensively profile compound LLM applications and identify uncertainty-reducing stages, along with an entropy-based mechanism to quantify their uncertainty reduction. Combining an uncertainty reduction strategy and a job completion time (JCT)-efficient scheme, we further propose an efficient scheduler to reduce the average JCT. Evaluation of both simulation and testbed experiments on various representative compound LLM applications shows that compared to existing state-of-the-art scheduling schemes, LLMSched can reduce the average JCT by 14$\sim$79\%.
\end{abstract}


\section{Introduction}
Large language models (LLMs) are revolutionizing the digital world and becoming the critical component of artificial general intelligence (AGI). Trained and fine-tuned on the massive human corpus, LLMs with billions of parameters demonstrate a remarkable ability to perform NLP tasks like content summarization, code generation, and machine translation commensurate with or beyond human experts. Notable examples of LLMs include powerful closed-source models like GPT-4 \cite{gpt4} and Gemini \cite{gemini}, and open-source models such as Llama \cite{llama}, Vicuna \cite{vicuna} and Qwen \cite{qwen}. Consequently, these models receive wide popularity and significant research attention from both industry and academia. For instance, GPT-4 \cite{gpt4}, a representative LLM nowadays, has about 180.5 million users with 100 million users active weekly \cite{gptstatistics}.

To handle more complex real-world problems, existing LLM applications have shifted from relying on a monolithic LLM to compound LLM systems, where an LLM collaborates with various external modules, including APIs and even other LLMs, to deliver complex intelligent services\cite{llmcompound} \cite{got} \cite{reflexion}. For example, in the task automation application \cite{taskbench}, the LLM acts as a planner, selecting deep learning models to solve complex tasks for users. To serve these applications, LLM service providers need to schedule a large volume of inference requests onto GPU-equipped clusters with a short response time. Compared with typical LLM requests, serving compound LLM applications is even more challenging as they can contain several or even tens of LLM inference tasks \cite{got} \cite{react}, along with other non-LLM tasks such as tool invocations.

For complex jobs with dependent stages that contain one or more parallel tasks for execution, it is a common practice to represent each job as a directed acyclic graph (DAG) where stages are nodes, and edges capture the input-output dependency. {In the context of compound LLM applications, they can similarly be modeled as DAGs. For example, a program synthesis application \cite{autogen} includes multiple iterative code executing and code generation stages, organized in a chain-like DAG.} Driven by the dependencies inherent in compound LLM applications, it is natural to consider leveraging the conventional cluster serving framework and applying existing DAG scheduling schemes to serve these applications.

However, based on our comprehensive experiments on three types of common compound LLM applications covering thousands of queries, we identify two key differences between compound LLM applications and existing data processing jobs such as database query \cite{decima} and regression analysis \cite{branch}. First, since the LLM generates the response autoregressively \cite{gpt4} \cite{orca}, the duration of LLM queries in compound LLM applications is highly uncertain and tricky to predict \cite{fastserve}. We find that the job\footnote{{In the context of compound LLM applications, we refer to a "job" as a runtime instance of an application with specific user input.}} duration of a compound LLM application can range from several to hundreds of seconds. The batch processing characteristic in LLM inference \cite{vllm} further intensifies the duration uncertainty. Second, a considerable part of compound LLM applications treats the LLM as an agent and asks it to plan \cite{taskbench} \cite{hugginggpt} or reason in an unfixed number of rounds \cite{reflexion}. This makes their structure non-deterministic. For example, the chain length in the code generation application \cite{reflexion} can vary from 3 to 15 steps. More details about these two types of uncertainty can be found in Sec.\ref{motivation}.

These two unique characteristics greatly challenge the scheduling of compound LLM applications. Specifically, existing scheduling schemes for DAG jobs typically rely on sufficient and accurate information about job duration and DAG structure \cite{decima} \cite{graphene} \cite{carbyne}. However, the highly uncertain job durations lead to inaccurate estimations, which in turn result in degraded performance for duration-based methods such as Shortest Job First (SJF) \cite{sjf}. In addition, the structural uncertainty of compound LLM applications makes the traditional DAG model inapplicable, as jobs from the same application can exhibit different and runtime-determined topologies. This restricts the use of topology-aware schedulers \cite{decima} \cite{argus} for emerging compound LLM applications.

To handle the above challenges, we propose LLMSched, a framework for efficiently scheduling online-arriving compound LLM applications at the provider's cluster. In LLMSched, we first design a novel DAG-based model to resolve the structural uncertainty in depicting compound LLM applications. Our model represents a compound LLM application as a DAG with \textit{regular stage}, \textit{LLM stage}, \textit{dynamic stage} and inter-stage dependencies. Aided by our proposed model, we design a Bayesian network-based profiler to comprehensively profile the duration and structure of compound LLM applications and identify uncertainty-reducing stages by mining the inter-stage correlation. {To efficiently mitigate the uncertainty,} we next design an entropy-based mechanism to quantify the uncertainty reduction of scheduling these stages. {This mechanism allows us to prioritize stages with greater quantified uncertainty reduction.} {Since our primary goal is to reduce the average JCT,}
we further propose an uncertainty-aware scheduler that combines an uncertainty reduction strategy and a JCT-efficient scheduling scheme using the $\epsilon$-greedy algorithm. For evaluating the performance of LLMSched at scale, we build a simulator to compare it with several SOTA schedulers and further implement a real-world testbed for practical testing. The main contributions of this paper are summarized as follows:
\begin{itemize}
    \item We conduct thorough data analysis on representative compound LLM applications. We identify the temporal and structural uncertainty of emerging compound LLM applications and reveal the challenges in scheduling. 
    \item We propose a novel DAG-based model for depicting compound LLM applications and a Bayesian network-based profiler to profile their durations and structures and identify uncertainty-reducing stages.
    \item We design an entropy-based uncertainty quantification mechanism and an uncertainty-aware scheduler by judiciously combining an uncertainty reduction strategy and a JCT-efficient scheme to reduce the average JCT.
    \item Experimental results from both large-scale simulations and a real-world testbed demonstrate that LLMSched can reduce the average JCT by up to 79\%  compared to the state-of-the-art scheduling schemes.
\end{itemize}

The rest of the paper is organized as follows. Section \ref{background} introduces the background of compound LLM applications, their scheduling problem and existing scheduling policies. Section \ref{motivation} presents the workload characterization and motivation for this work. Section \ref{framework section} introduces the LLMSched framework in detail. Section \ref{experiment} presents the experimental results. Section \ref{relatedwork} discusses the related works. Section \ref{conclusion} concludes the paper.

    \begin{figure*}
        \centering
        \subfloat[Uncertainty in job duration.]{
        \includegraphics[width=0.30\textwidth]{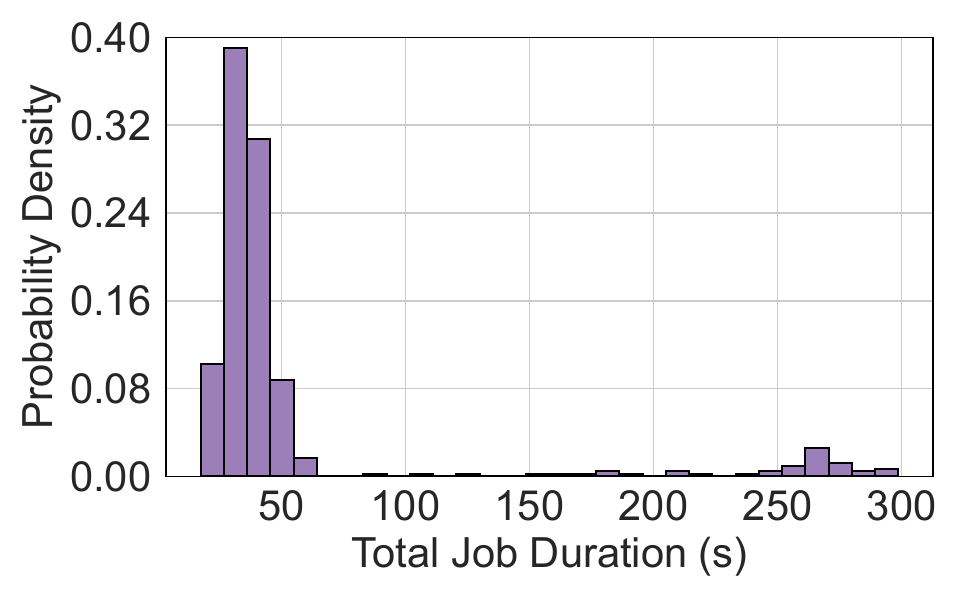}
            \label{sorting_duration}
        }
        \subfloat[Structural uncertainty in chain length.]{
            \includegraphics[width=0.30\textwidth]{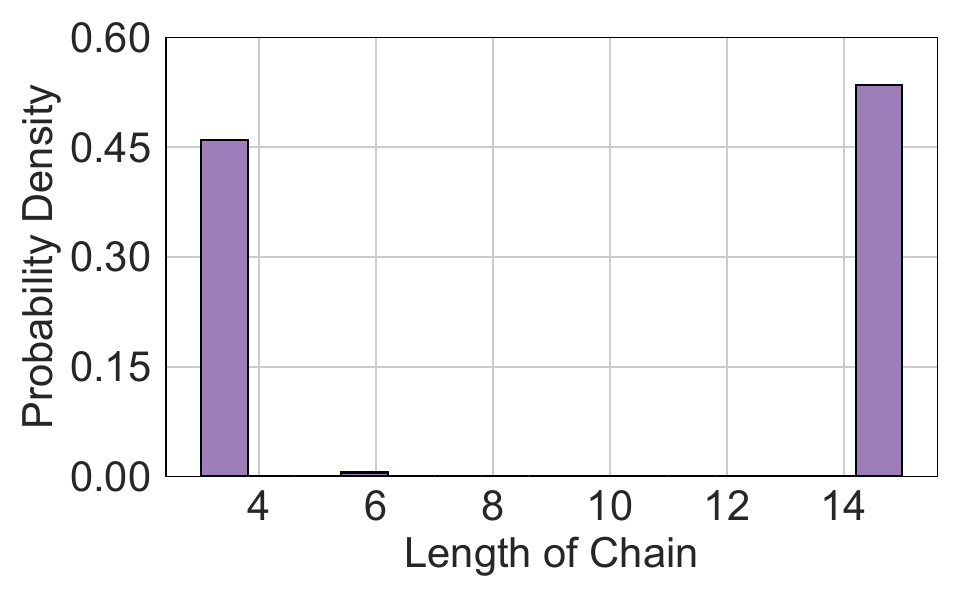}
            \label{codegenlength}
        }
           \subfloat[Structural uncertainty in generated stages.]{
            \includegraphics[width=0.30\textwidth]{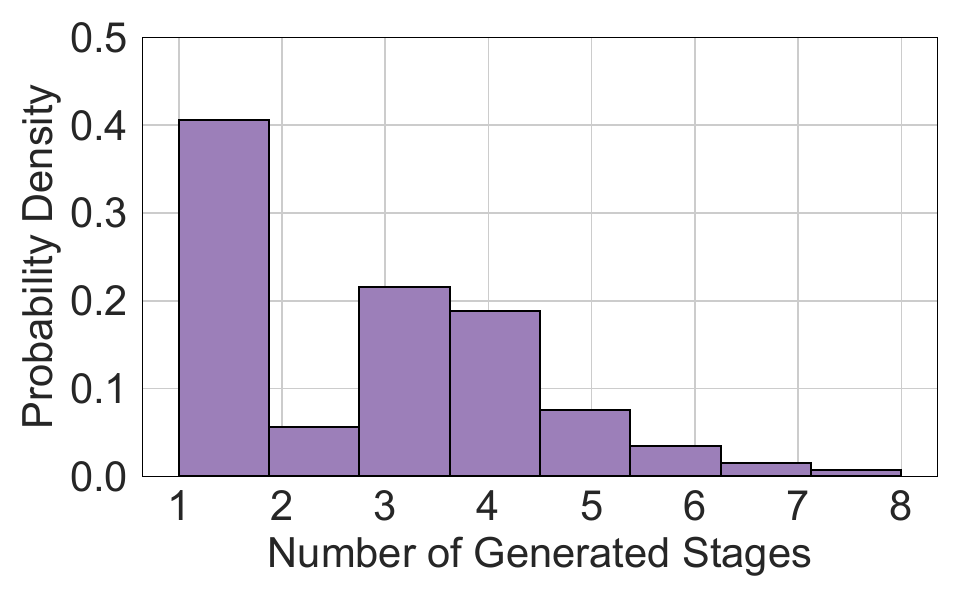}
            \label{hugginggptnumstage}
        }
        \caption{Runtime characteristics of three representative compound LLM applications: (a) sequence sorting; (b) code generation; (c) task automation.}
        \label{appruntime}
    \end{figure*}

    \begin{figure}

        \centering
        \subfloat[Two jobs arrive at 0s. Job 1 is a job of task automation application and Job 2 is a job of code generation application.]{
        \includegraphics[width=0.47\textwidth]{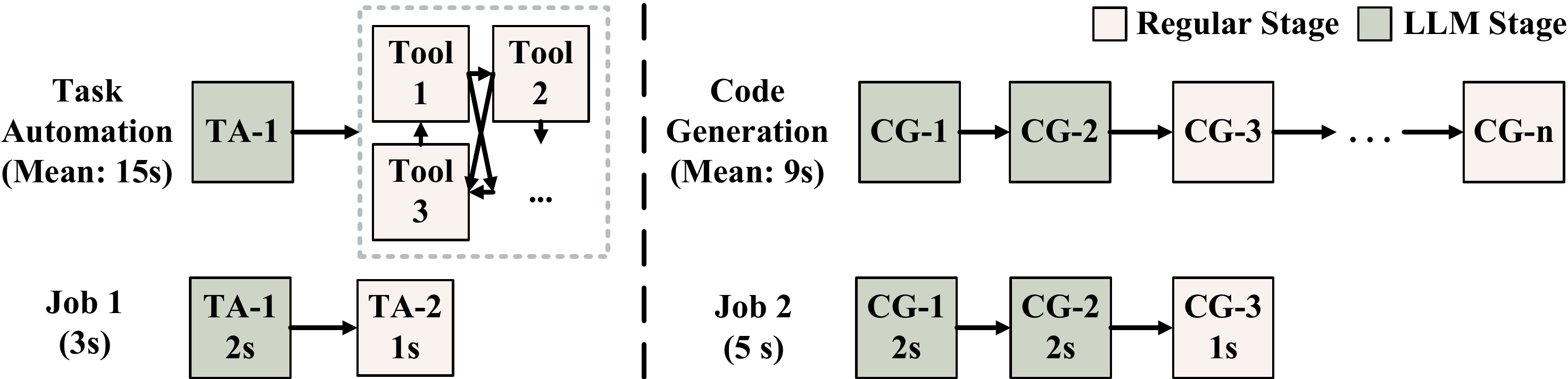}
            \label{jobexample}
        }\\
        \subfloat[Shortest Job First scheduling method finishes two jobs with an average JCT of 6.5s.]{
            \includegraphics[width=0.22\textwidth]{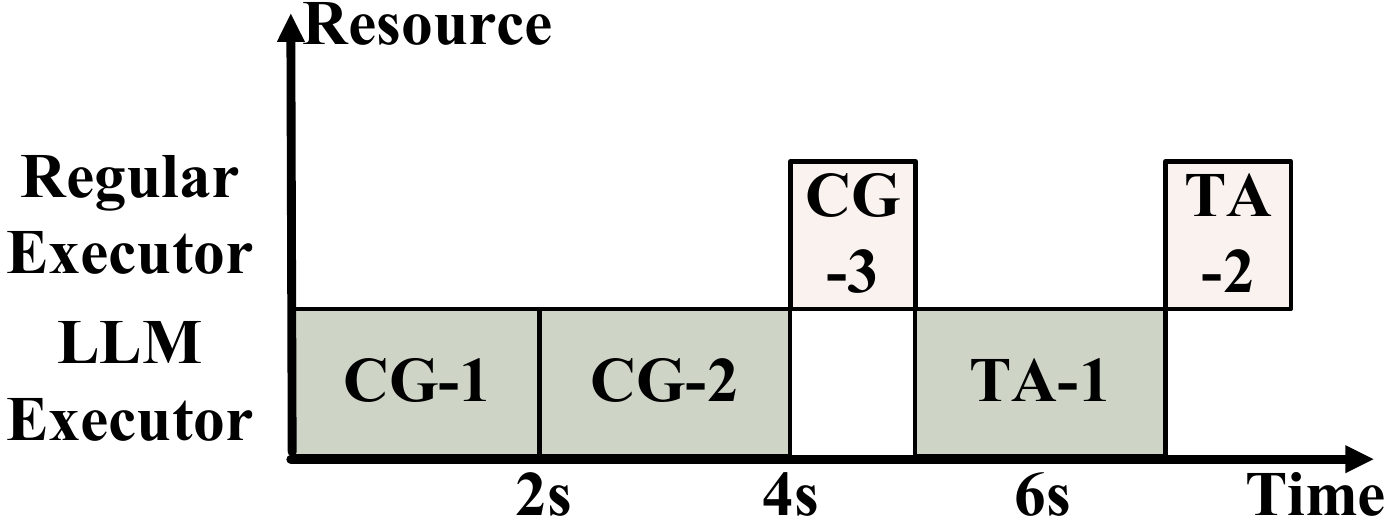}
            \label{SJFexample}
        } \hspace{0.001cm}
        \subfloat[Uncertainty-aware scheduling method finishes two jobs with an average JCT of 5s.]{
            \includegraphics[width=0.22\textwidth]{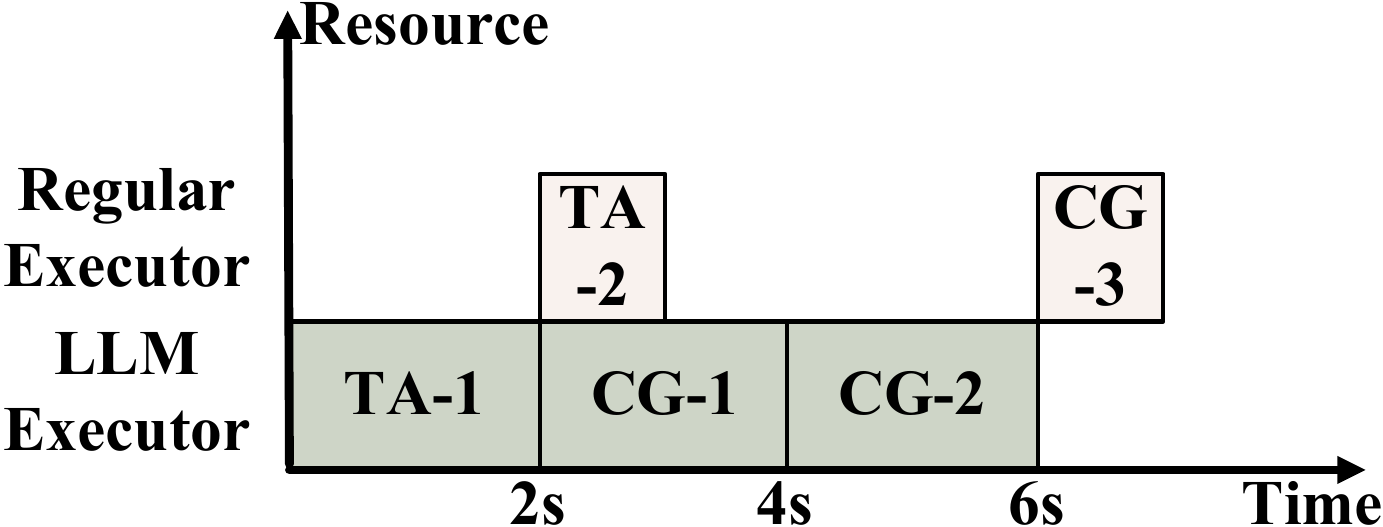}
            \label{Uncertainty-aware example}
        }
        \caption{An example of the benefits of considering uncertainty for scheduling compound LLM applications.}
        \label{motivationexample}
    \end{figure}  

\section{Background}
\label{background}
\subsection{Compound LLM Applications}
We conduct a comprehensive review of existing compound LLM applications, thoroughly analyzing their design principles, execution workflows, and practical implementations.
Based on this analysis, we classify them into three types:
\begin{itemize}
    \item \textbf{Predefined applications} \cite{got} \cite{langchain}  have an explicitly defined workflow with a fixed number of stages and dependencies similar to traditional data processing jobs \cite{spark} \cite{hadoop}. A representative example is the sequence sorting in Graph-of-Thoughts \cite{got}. In sequence sorting, the LLM first divides the original sequence into multiple parts. For each part, the LLM generates multiple candidates, along with user-defined functions to score and select candidates. The LLM then merges and processes the selected candidates as the final sorting results.

    \item \textbf{Chain-like applications} \cite{reflexion} \cite{react} operate step-by-step and follow an iterative operation pattern. A representative example is the code generation \cite{reflexion} \cite{codebench}. Given a user's query, code generation first uses the LLM to generate several use cases for code testing and then iteratively generates code (LLM), executes the code, and reflects on the results (LLM). The iteration continues until the generated code passes the test cases or the maximum number of iterations is reached.
    
    \item \textbf{Planning applications} \cite{hugginggpt} \cite{llmcompiler} request the LLM to generate a plan (e.g., a set of stages for invoking external modules or LLM and their dependencies) and execute it to fulfill the user's query. A representative example is the task automation \cite{hugginggpt}, where the LLM analyzes the user's task, selects tools like deep learning models, and identifies dependencies between the selected tools.
\end{itemize}
\subsection{Scheduling Problem in Compound LLM Applications}
In this paper, we focus on the scheduling problem for compound LLM applications on a cluster of machines with GPUs to reduce response time. Over a certain period, the jobs of different compound LLM applications arrive continuously. Each job, in hindsight, can have multiple stages with inter-stage dependencies. The exact stages and their dependencies may be unknown at the beginning based on different compound LLM applications. Depending on whether a stage is realized by an LLM, we divide stages into LLM stages or regular stages, and their tasks into LLM tasks and regular tasks, respectively. The LLM service provider deploys regular executors (e.g., containers) \cite{decima} \cite{spark} and LLM executors (e.g., LLM engine) with a specified maximum batch size \cite{vllm} to serve the jobs. Each regular executor can execute one regular task at a time and each LLM executor can execute multiple LLM tasks concurrently, up to its maximum batch size. Our goal is to schedule the arrived jobs of compound LLM applications to minimize the average JCT.
\subsection{Existing Scheduling Policies and Their Deficiency.}
 To schedule compound LLM applications, we can utilize either job-agnostic methods \cite{spark} or existing DAG scheduling schemes \cite{decima} \cite{carbyne} \cite{argus} \cite{spark} considering their inherent inter-stage dependencies. Job-agnostic methods such as First Come First Serve and Fair Scheduling \cite{spark}, prioritize jobs based on arrival time or fairness principles without considering job information. Duration-based methods like Shortest Job First \cite{sjf} and Carbyne \cite{carbyne} schedule jobs according to historical job duration. Topology-aware methods such as Decima \cite{decima} and Argus \cite{argus} incorporate DAG topology of jobs for efficient scheduling.
 
 However, job-agnostic methods are generally suboptimal for reducing average JCT. Current DAG scheduling schemes typically rely on sufficient and accurate information (e.g., duration and DAG structure) about jobs, which proves to be deficient when scheduling compound LLM applications \cite{decima}. The inherent duration uncertainty in compound LLM applications makes duration estimation inaccurate, which degrades the performance of duration-based methods. Additionally, the capability of topology-aware methods is also restricted due to the structural uncertainty in compound LLM applications. \textit{These limitations in current scheduling schemes motivate us to design a new scheduler tailored for compound LLM applications.}

\section{Motivation Exploration}
\label{motivation}
\subsection{Workload Characterization}

In this part, we analyze the runtime characteristics of compound LLM applications. We choose three representative compound LLM applications mentioned in the previous section and evaluate them on different datasets. For predefined applications, we select the sequence sorting application and run it on a synthetic dataset with 500 randomly generated sequences of lengths ranging from 16 to 64. For chain-like applications, we select the code generation application and run it on the MBPP dataset \cite{mbpp}, which contains 974 programming tasks and is commonly used to evaluate the code generation performance of LLMs. For planning applications, we select the task automation application and run it on TaskBench \cite{taskbench}, a benchmark for evaluating LLMs' task automation ability with tens of tools and thousands of queries. The experiments are conducted on a server with an H800 GPU, using the Llama 2-7B model \cite{llama} with a batch size of 1. We present our main observations as follows:\\
\textbf{Observation 1: The duration of a job is uncertain.}
Fig. \ref{sorting_duration} shows the distribution of the job duration for sequence sorting. We find that the job duration varies widely, ranging from 10s to nearly 300s. This is quite different from traditional data processing applications whose job durations are typically stable and certain \cite{decima} \cite{carbyne}. Code generation and task automation also show large uncertainty in job duration, which ranges from 2s to 50s and 1s to 116s, respectively. We omit visualizing them here due to space limitations. Our findings extend previous observations on the duration of single LLM requests \cite{fastserve} to compound LLM applications, confirming their temporal uncertainty.\\
\textbf{Observation 2: The structure of a job is uncertain.}

We now examine the number of stages in code generation and task automation and plot the distribution of the number of stages. Code generation follows a chain-like workflow based on code execution results, with the chain length being dynamic. As shown in Fig. \ref{codegenlength}, the chain length ranges from 3 to 15 stages. Similarly, we find that the number of generated stages in task automation is also dynamic, ranging from 1 to 8, as shown in Fig. \ref{hugginggptnumstage}. Note that the number of stages of both applications is not predefined, but rather revealed during the execution process. These results reveal the structural uncertainty of compound LLM applications.
\subsection{Motivation for Uncertainty-aware Scheduling}
    We use a simple illustrative example to demonstrate how the uncertainty of compound LLM applications influences the performance of traditional DAG scheduling schemes. Let's consider a simple scheduling scenario shown in Fig. \ref{motivationexample}. Here, the green block and the beige block represent the LLM stage and the regular stage, respectively. The left part of Fig. \ref{jobexample} shows the general workflow of task automation with an average duration of 15s (top), and an actual job of this application, referred to as Job 1, with its actual structure and duration of 3s (bottom). The right part of Fig. \ref{jobexample} shows the general workflow of code generation with an average duration of 9s (top), and an actual job, referred to as Job 2, with its actual structure and duration of 5s (bottom). Each stage is annotated with its name and duration. For example, "TA-1 2s" indicates that the stage named "TA-1" of Job 1 will take 2s to complete, and "CG-1 2s" indicates that the stage named "CG-1" of Job 2 will take 2s to complete. We set the number of LLM executors, the maximum batch size, and the number of regular executors to 1. Both jobs arrive at 0s. 

The Shortest Job First (SJF) method, which is known to be efficient for minimizing average JCT \cite{carbyne}, schedules jobs based on their estimated duration. Thus, as shown in Fig. \ref{SJFexample}, it first schedules Job 2 and then Job 1 based on the expected duration of each job. Unfortunately, due to the uncertainty of compound LLM applications, the actual duration of Job 2 is nearly twice that of Job 1, which is contrary to the estimation and results in a high average JCT of 6.5s. 
    \begin{figure}[]
        \centering
        \includegraphics[width=0.48\textwidth]{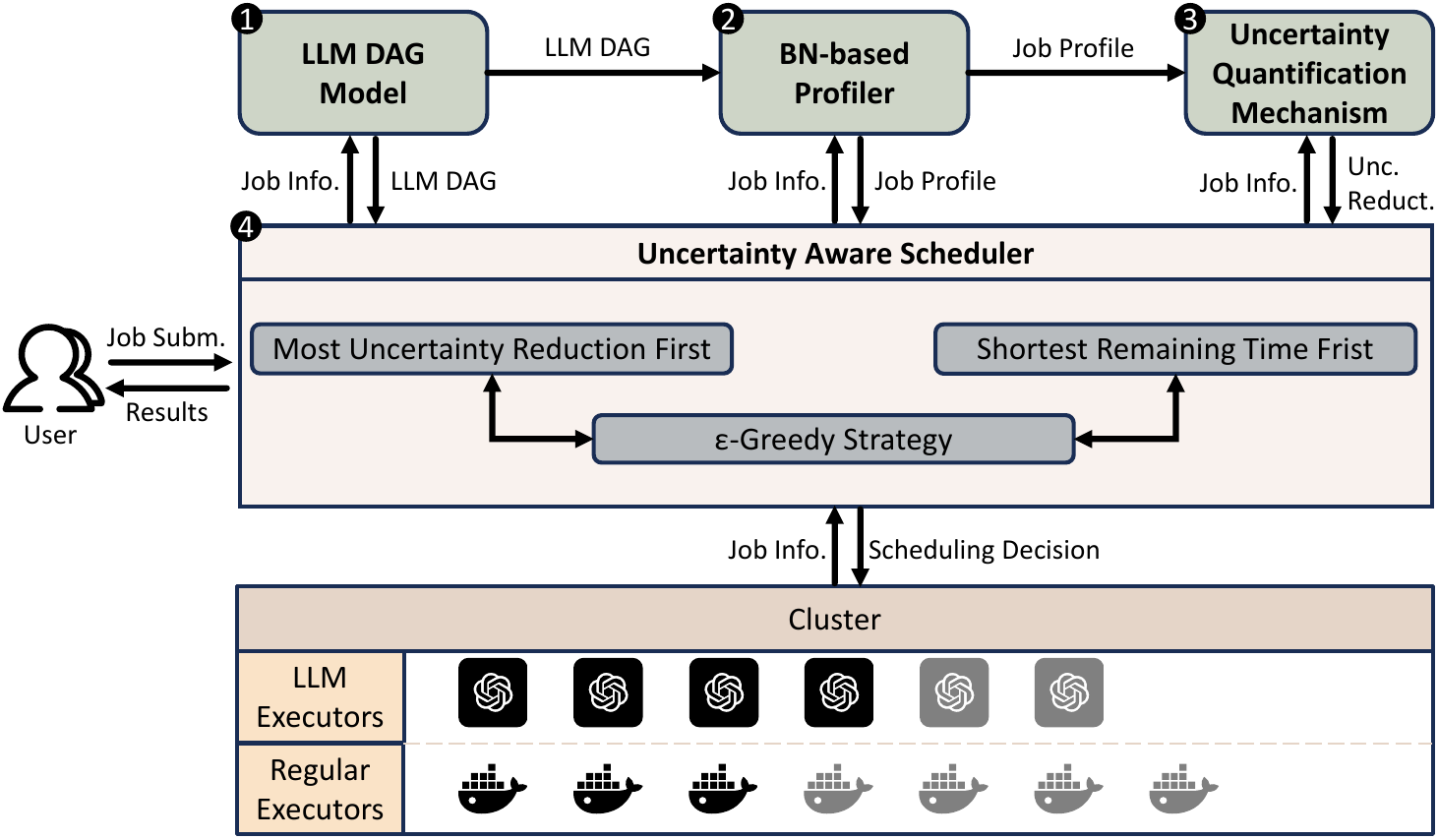}
        \caption{LLMSched Overview}
        \label{overview}
    \end{figure}  

    The SJF method overlooks the inherent uncertainty in compound LLM applications, leading to suboptimal performance. \textit{In contrast, we propose that by first focusing on scheduling stages that reduce uncertainty, we can obtain valuable information once they're completed. Such information can be leveraged to improve the overall performance of serving all jobs of compound LLM applications.} As illustrated in Fig. \ref{Uncertainty-aware example}, we identify that stage TA-1 has a strong correlation with the remaining undetermined stages since the rest of Job 1 is entirely dependent on TA-1's LLM output. Once we finish TA-1, its uncertainty in both duration and structure is fully resolved. Consequently, we first schedule TA-1 and find that Job 1's duration is much shorter than the duration expectation of Job 2. Thus, we then apply the Shortest Remaining Time First (SRTF) to schedule the remaining stages. This uncertainty-aware scheduling principle improves the average JCT to 5s.

    This simple example demonstrates the great significance of uncertainty awareness, i.e., identifying and prioritizing uncertainty-reducing stages, for scheduling compound LLM applications. However, to incorporate uncertainty awareness for scheduling, we need to answer the following questions: \textit{1) How to identify the stages that can reduce uncertainty after being scheduled? 2) If there are many uncertainty-reducing stages, how to decide which stage should be scheduled first? 3) How to leverage uncertainty awareness to achieve our goal, i.e., reducing the average JCT?} To address these three questions, we next introduce our LLMSched framework.

\section{LLMSched Framework}
\label{framework section}
In this section, we present LLMSched, an efficient scheduling framework for compound LLM applications. As shown in Fig. \ref{overview}, LLMSched contains four main components: First, a novel DAG-based model for depicting compound LLM application. Second, a Bayesian network-based profiler to profile the compound LLM applications and identify uncertainty-reducing stages. Third, an entropy-based mechanism for uncertainty quantification. Fourth, an uncertainty-aware scheduler for efficient scheduling.

\subsection{DAG-based Model for Compound LLM Applications}
In this section, we introduce our DAG-based model for depicting compound LLM applications, which serves as the basis for solving the three questions in Sec.\ref{motivation}. Due to the structural uncertainty of compound LLM applications and the unique characteristics of LLM inference tasks, the traditional DAG format does not apply. Therefore, we design a novel DAG-based model to resolve the structural uncertainty for depicting these applications. Our model characterizes existing compound LLM applications using three types of stages, and treats stages of each type as nodes and the inter-stage dependencies as edges. The three types of stages are as follows:

\begin{itemize}
    \item \textbf{Regular stage}: A \textit{regular stage} is a stage that contains one or more regular (i.e., non-LLM) tasks and is identical to the stage definition in the traditional DAG model \cite{spark}. The regular tasks will be executed on regular executors such as containers \cite{llmcompiler}.
    \item \textbf{LLM stage}: An \textit{LLM stage} is a stage that contains one or more LLM inference tasks. The LLM tasks will be executed on the LLM executors, with each executor running an LLM instance \cite{vllm}.
    \item \textbf{Dynamic stage}: A \textit{dynamic stage} acts as a placeholder for those LLM-generated stages and their dependencies, which are constructed by the preceding \textit{LLM stage} from a candidate set. The candidate set includes other regular stages or LLM stages for the LLM to invoke.
\end{itemize}
    \begin{figure}[]
        \centering
        \includegraphics[width=0.49\textwidth]{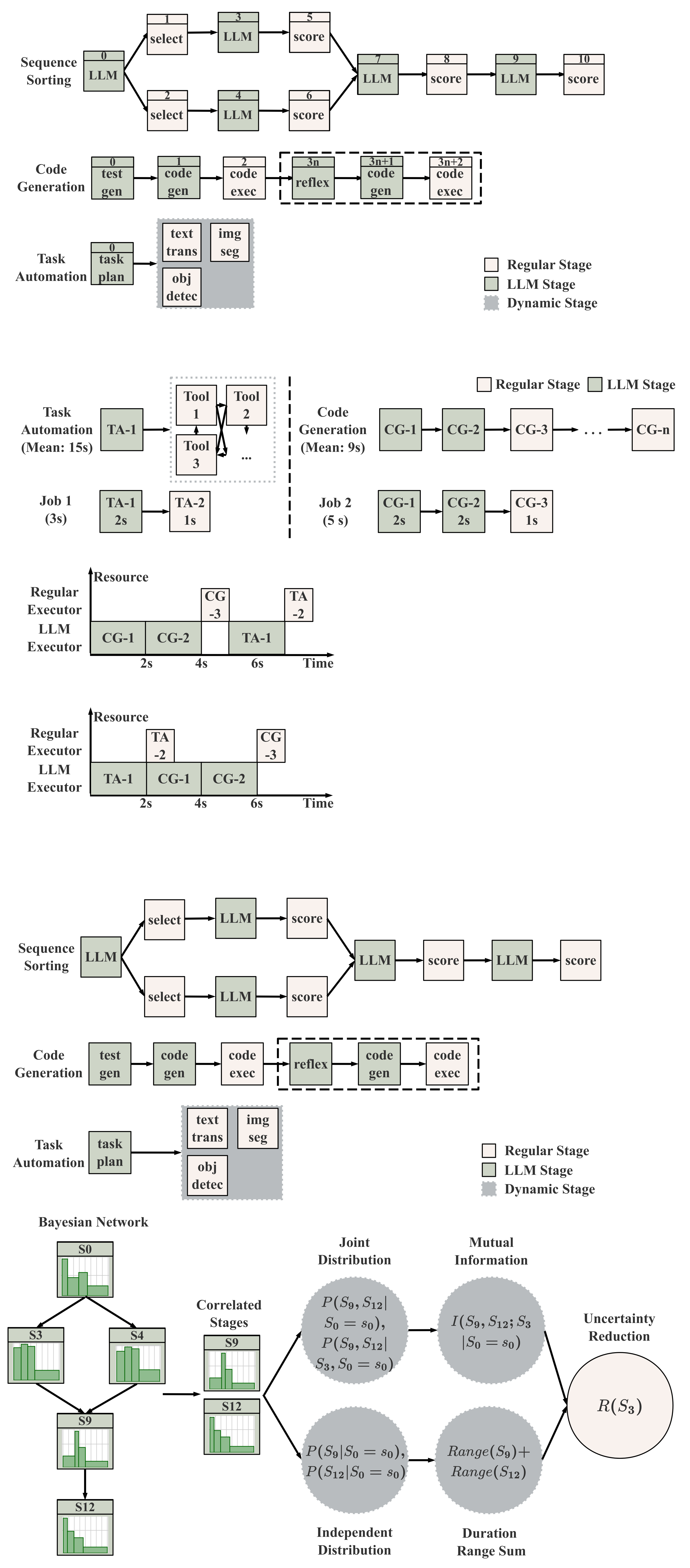}
        \caption{LLM DAG representations for sequence sorting, code generation, and task automation.}
        \label{dagexample}
    \end{figure}

As discussed in Sec.\ref{motivation}, the structural uncertainty exists in chain-like and planning applications. For current chain-like applications, we notice that they typically iterate over a fixed pattern with a specified maximum number of iterations. Therefore, we can pad the chain based on the pattern and the maximum number of iterations to resolve their structural uncertainty for depiction. For planning applications, the design of the \textit{dynamic stage} directly abstracts away their uncertainty.

    \textbf{Illustrative Examples:}
    In Fig. \ref{dagexample}, we illustrate how to depict a compound LLM application as a DAG using our model. We provide depictions of three applications mentioned in Sec. \ref{motivation}. {The number of each stage represents its stage ID sorted in a topological order.} The first application, sequence sorting, is predefined with explicit dependencies, so we use the structure defined in the original paper \cite{got}. The second example is the code generation application. Due to space limitations, we only illustrate the depiction with a maximum of one iteration. The stages in one iteration are circled with a black dashed rectangle. For the task automation application, we use an \textit{LLM stage} for task planning and a \textit{dynamic stage} with a stage candidate set including three \textit{regular stages} (text translation, image segmentation and object detection) to represent the generated plan.

\subsection{Bayesian Network-based Profiler}
\label{bayesian section}
\begin{figure}
        \centering
        \subfloat[Sequence sorting application.]{
        \includegraphics[width=0.23\textwidth]{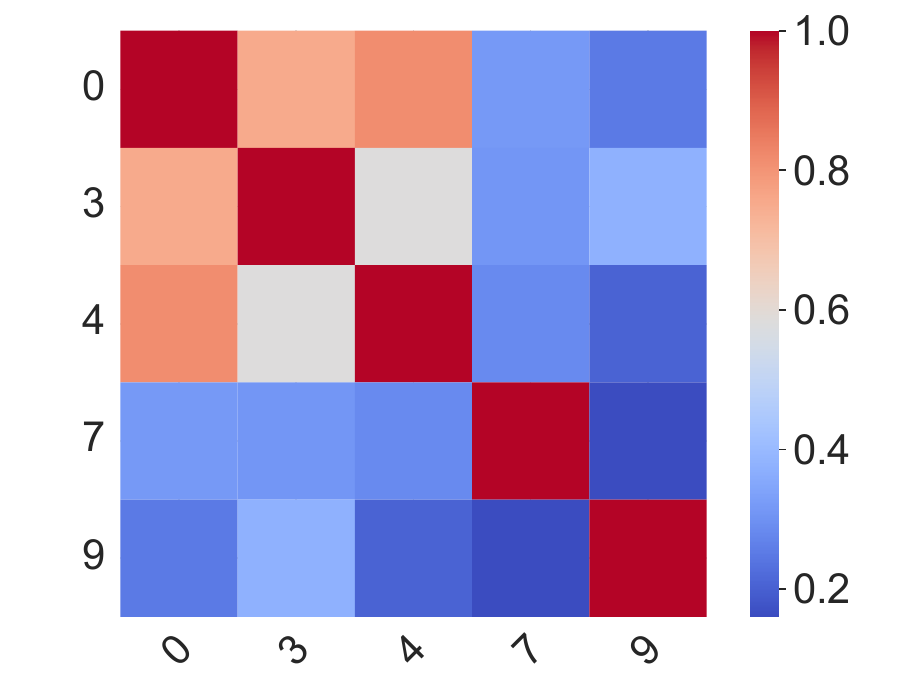}
            \label{sortingheapmap}
        }
        \subfloat[Code generation application.]{
            \includegraphics[width=0.23\textwidth]{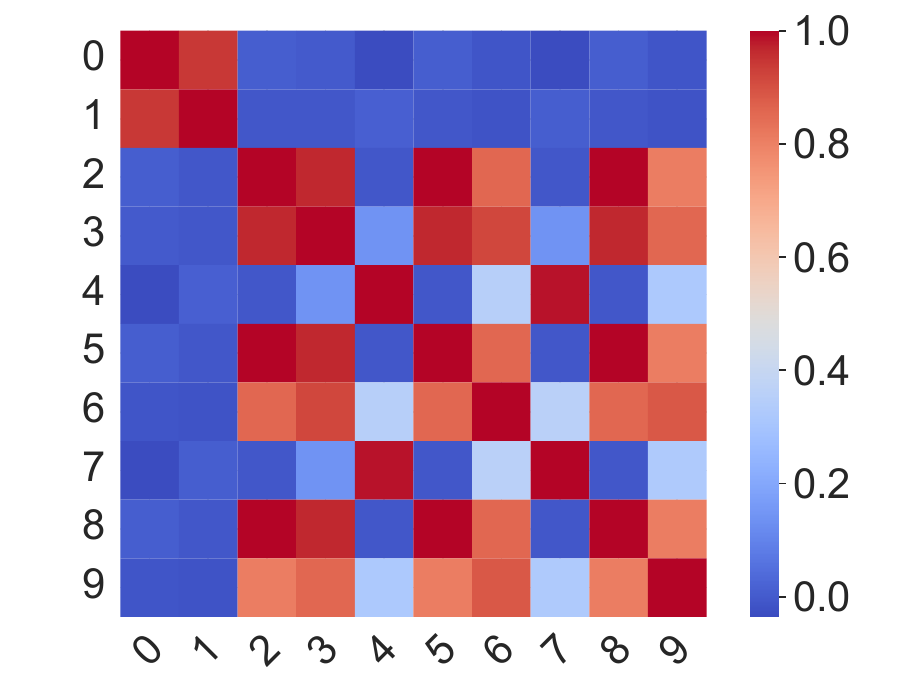}
            \label{codegenheatmap}
        }
        \caption{Heatmap for the duration of the stages in two compound LLM applications. {The axis records the stage IDs sorted in a topological order with respect to the DAG in Fig. \ref{dagexample}.}}
        \label{heatmap}
    \end{figure}

In this section, we answer the first question: How to identify the stages that can reduce uncertainty after being scheduled? The motivation example in Sec.\ref{motivation} shows the uncertainty reduction resulting from scheduling the correlated \textit{LLM stage} of the \textit{dynamic stage}. Inspired by this, we wonder if we can leverage the inter-stage correlations to identify uncertainty-reducing stages. The correlation between the \textit{dynamic stage} and its preceding \textit{LLM stage} is clear in planning applications, as the stages and dependencies in the \textit{dynamic stage} are determined by the output of the \textit{LLM stage}. Therefore, we further investigate the correlation within and between \textit{regular stages} and \textit{LLM stages} in predefined and chain-like applications. To this end, we plot heatmaps showing the duration of different stages of two representative applications: sequence sorting (predefined) and code generation\footnote{Since different jobs of code generation have different numbers of iterations, we treat the duration of those unexecuted stages as 0s at runtime.} (chain-like). The heatmaps use Pearson coefficients, where a value approaching 1 or -1 indicates a positive or negative linear correlation, respectively. As shown in Fig. \ref{heatmap}, some stages exhibit strong correlations in their durations. For example, stages 0 and 3 of sequence sorting have a coefficient of 0.7, and stages 3 and 6 of code generation have a coefficient of nearly 0.9.

The strong correlation between the durations of different stages reflects the compound LLM application's intrinsic design logic. In sequence sorting, the LLM splits the original sequence into two equal parts at stage 0 and processes these parts at stages 3 and 4 to produce sorted results. These stages exhibit high correlation coefficients due to the proportional relationship in the number of generated tokens. The correlations in code generation stem from two factors: first, each \textit{code gen} stage shown in Fig. \ref{dagexample} modifies the previously generated code based on execution results, leading to similar code and duration; second, the iterative nature ensures all stages in a standard iteration are executed sequentially. For instance, a \textit{reflex} stage implies subsequent \textit{code gen} and \textit{code exec} stages.

Traditional scheduling schemes \cite{decima} \cite{carbyne} adopt single-job or single-stage duration profiling, missing the opportunity to leverage inter-stage correlations for uncertainty reduction. To comprehensively profile compound LLM applications and capture such correlations for identifying uncertainty-reducing stages, we leverage the Bayesian network (BN) \cite{bayesiannetwork}. A BN is a probabilistic graphical model representing a set of variables and their conditional dependencies using a DAG. By discretizing the duration distribution of each stage in a compound LLM application, we can train a BN to derive the conditional probability distributions (CPDs) of stages' durations and their correlations. CPDs specify the probability distribution of a variable given its parent variables in the network, allowing us to model the duration of each stage depending on the durations of preceding stages. The edges in the BN further represent these dependencies, revealing the inter-stage correlations and identifying uncertainty-reducing stages. {Specifically, given two stages $u$ and $v$ in the BN, $u$ is considered \textit{correlated with} $v$ if a directpath exists between them in the BN:
\begin{align}
\label{correlated}
    \text{correlated}(u, v) = 
\begin{cases}
1 & \text{if } u \overset{*}{\to} v, \\
0 & \text{otherwise}.
\end{cases}
\end{align}
We treat all stages correlated with one or more other stages as uncertainty-reducing.}

With the BN, we can also dynamically update the profile of job duration and structure for more accurate estimation. This is achieved by inferring the posterior distribution of unfinished stages using the durations of completed stages and the obtained CPDs. For instance, if early stages take longer than expected, the BN can adjust the estimated durations of subsequent stages accordingly. This inference capability of the BN enables real-time adjustments and more precise scheduling decisions, thus reducing uncertainty and improving efficiency.

\begin{figure}[]
    \centering
    \includegraphics[width=0.49\textwidth]{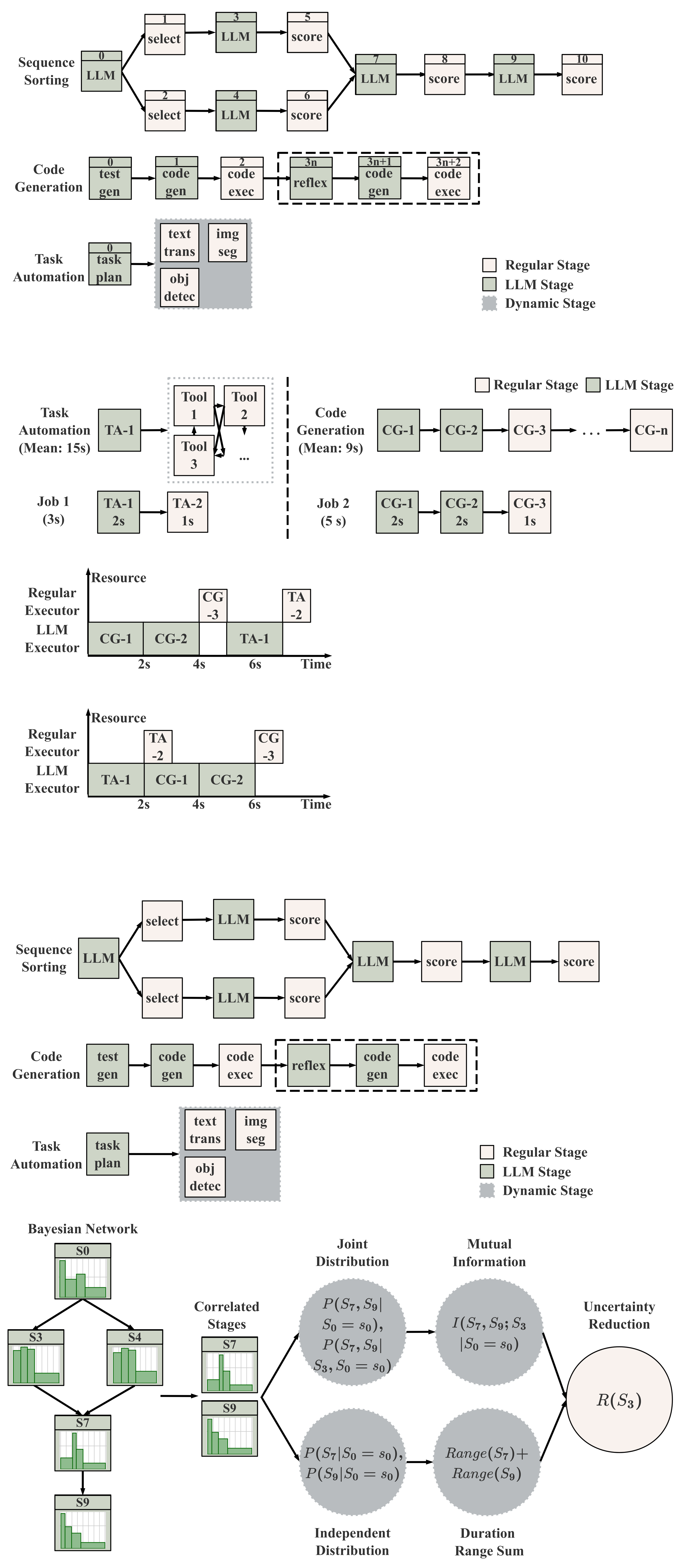}
    \caption{{An example for calculating the uncertainty reduction of scheduling $S_3$ in sequence sorting application.}}
    \label{example_uncer_reduct}
\end{figure} 

\textbf{Batching-aware Duration Calibration:} As mentioned earlier, the duration of an LLM task is affected by the number of runtime batched requests, which intensifies the duration uncertainty and can result in inaccurate duration estimation. Thus, we propose batching-aware duration calibration, which adjusts the estimation of $LLM stage$ duration based on current batched requests, to aid the BN in runtime job duration estimation. We simulate workloads with different batch sizes and record the average latency for decoding one token. For an LLM task with duration $d_r$ executed under batch size $b_r$, we calibrate the duration estimation $d_t$ at the target batch size $b_t$ as follows:
\begin{align}
    d_t = d_r \times \frac{l(b_t)}{l(b_r)},
    \label{calibration}
\end{align}
where $l(b_t)$ and $l(b_r)$ are the recorded average decoding latencies under the corresponding batch sizes.

\subsection{Entropy-based Uncertainty Quantification}
In this section, we answer the second question: If there are many uncertainty-reducing stages, how to decide which stage should be scheduled first? The CPDs obtained by the BN inspire us to leverage Shannon entropy \cite{entropy}, a distribution-sensitive metric for measuring the uncertainty of random variables, to quantify the uncertainty reduction of scheduling a stage. Given a discrete random variable $X$ with $n$ possible values, the entropy $H(X)$ is defined as:
\begin{align}
    H(X) = -\sum_i^n p(x_i) \log_2 p(x_i),
    \label{shannon entropy}
\end{align}
where $p(x_i)$ is the probability of $X$ taking the value $x_i$. 

{To leverage the Shannon entropy for uncertainty quantification,} we first design random variables to characterize the three types of stages in our DAG model as follows:
\begin{itemize}
    \item \textbf{The uncertainty of a \textit{regular stage}} lies in whether that stage will be executed or not since the duration of a regular task is generally certain \cite{decima} \cite{graphene}. We can simply use a binary random variable to represent its execution.
    \item \textbf{The uncertainty of an \textit{LLM stage}} lies in both its execution and its duration.  
    By discretizing the duration distribution into $k$ intervals and treating the non-execution as duration 0s, we can create a random variable with $k + 1$ distinct values to represent the \textit{LLM stage}.
    \item \textbf{The uncertainty of a \textit{dynamic stage}} lies in its inner stages and dependencies, which can be characterized by the selection of stages in the candidate set and edges. We define the uncertainty of a \textit{dynamic stage} as the sum of node (stage) entropy and edge (dependency) entropy \cite{dagentropy}, and formulate it as follows:
        \begin{align}
            H(X) = \sum_{c \in \mathcal{C}} H(I_c) + \sum_{e \in \mathcal{E}} H(I_e),
            \label{dynamicstageuncertainty}
        \end{align}
        where $\mathcal{C}$ is the stage candidate set, $\mathcal{E}$ is a set of possible edges between the stage candidates in $C$, and $I_c$ and $I_e$ are binary variables that indicate the existence of stage $c$ and edge $e$, respectively.
\end{itemize}

{Next, we illustrate how we quantify the uncertainty reduction of scheduling a stage $X$. Let $Y$ denote an unscheduled stage correlated with $X$, we first introduce the mutual information $I(Y;X)$, a measure of the extent to which knowledge of $X$ reduces entropy about $Y$:
\begin{align}
    I(Y;X) &= -\sum_{x,y} p(x,y) \log_2 \frac{p(x,y)}{p(x)p(y)}
    \label{mutal information}
\end{align}
where $x$ and $y$ are the possible discretized duration values (e.g., duration interval) of $X$ and $Y$, respectively.} To incorporate the variation of all correlated stages' durations, we further refine the entropy reduction by considering the distribution range sum of all the correlated stages to calculate the final uncertainty reduction:
\begin{align}
    R(X) = I(Y_1,...,Y_M; X | \mathbf{E}) \times \sum_{m=1}^{M}\text{Range}(Y_m),
    \label{final uncertainty reduction}
\end{align}
where $Y_1, ..., Y_M$ are the unscheduled stages \textit{correlated with} $X$, and $\mathbf{E}$ is a set of evidence for the duration of all completed stages. For the \textit{dynamic stage}, since its uncertainty will be resolved when its preceding \textit{LLM stage} is finished, we multiply the uncertainty of the \textit{dynamic stage} measured using Eq. \eqref{dynamicstageuncertainty} by its duration distribution range, and add the result to its preceding \textit{LLM stage}'s uncertainty reduction.

	\begin{algorithm}[!t]
    \caption{Uncertainty-aware Scheduling}
    \label{schedulingalgorithm}
        \begin{algorithmic}[1]
        \Require A list of unfinished LLM jobs $J$,  exploration probability $\epsilon \in [0,1]$, task sampling ratio $r\in [0,1]$.
        \Ensure Scheduling preference for schedulable LLM tasks and regular tasks.
        \State $J_t$ = $sort(J, by = job.est\_rd())$
        \ForAll{$job$ in $J_t$ }
            \State $S_t.append(job.ready\_stages())$
        \EndFor
        \State Find the non-overlapping sets $J_{no}$ of $J$. 
        \ForAll{$j_{no}$ in $J_{no}$ }
            \State Get schedulable stages of each job in $j_{no}$ as $s_{no}$.
            \State $s'_{no} = sort(s_{no}, by = stage.uncert\_reduct())$
            \State $S_u.extend(s'_{no})$
        \EndFor
        \While {$S_t.has\_stage() \ and \ S_u.has\_stage()$}
            \State $s_t = S_t.pop()$, $s_u = S_u.pop()$            
            \State $p$ = $random(0,1)$
            \If{$p \leq \epsilon$}
                \State $t = s_u.sample\_tasks(r)$
            \Else
                \State $t = s_t.all\_tasks()$
            \EndIf
            \State Attach $t$ to $T_l$ if $t$ are LLM tasks otherwise $T_r$.
        \EndWhile
        \State Attach the remaining tasks to $T_l$ or $T_r$.
        \State Return $T_r$, $T_l$.
    \end{algorithmic}
\end{algorithm}

{\textbf{Illustrative Example:} We provide an example of sequence sorting application in Fig. \ref{example_uncer_reduct} to present how to calculate the uncertainty reduction of scheduling a stage. The left side of the figure shows the learned BN of the sequence sorting application. Assume we've completed $S_0$ and obtained its execution duration as $s_0$, and we want to calculate the uncertainty reduction resulting from scheduling $S_3$. We first gather all the correlated stages of $S_3$, namely $S_7$ and $S_9$ according to Eq. \eqref{correlated}. According to  Eq. \eqref{mutal information} and \eqref{final uncertainty reduction}, we obtain the two joint duration distributions, namely $P(S_7, S_{9} | S_0 = s_0)$ and $P(S_7, S_{9} | S_3, S_0 = s_0)$ from the BN, and use them to calculate the mutual information between stage $S_7$, $S_9$ and $S_3$. We next obtain the independent duration distributions of each correlated stage, namely $P(S_7 | S_0 = s_0)$ and $P(S_{9} | S_0 = s_0)$, and calculate the distribution range sum. At last, we multiply the mutual information and the distribution range sum to obtain the final uncertainty reduction of scheduling $S_3$.}

\begin{figure*}
    \centering
    \includegraphics[width=1\textwidth]{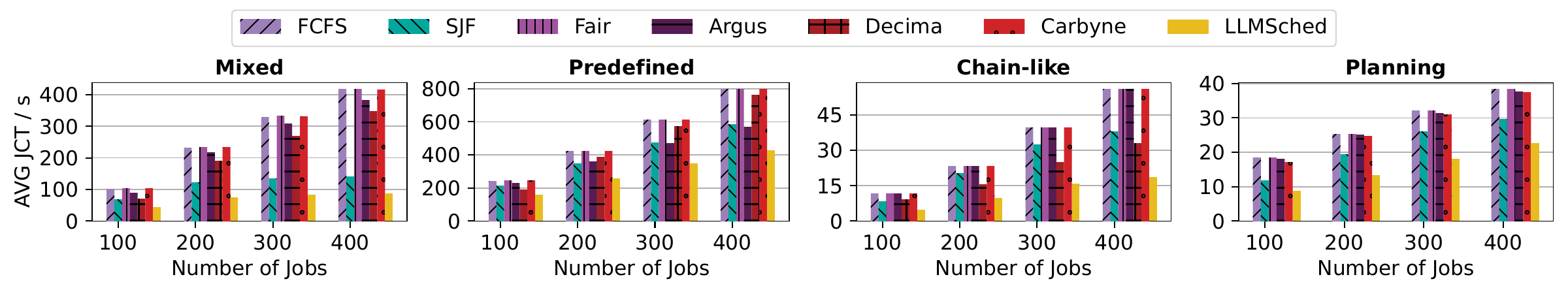}
    \caption{Simulation result for four types of workload ($\lambda = 0.9$). Decima is omitted in the Planning workload because its average JCT is too high (more than 100s).}
    \label{simulation result}
\end{figure*}

\subsection{Uncertainty-aware Scheduler}
In this section, we answer the final question: How to leverage uncertainty awareness to achieve our goal, i.e., reducing the average JCT? Our idea is to consider the process of reducing job uncertainty as exploration, and the use of obtained job information to identify and schedule small jobs as exploitation. Following this, we propose our uncertainty-aware scheduler that integrates an uncertainty reduction strategy with a JCT-efficient scheduling scheme, namely Shortest Remaining Time First (SRTF), in an exploration-exploitation manner. 

The details are shown in Algorithm \ref{schedulingalgorithm}, which takes a list of unfinished jobs $J$, an exploration probability $\epsilon$ and a task sampling ratio $r$, and outputs two scheduling preference lists: $T_r$ for regular tasks and $T_l$ for LLM tasks. First, we use the JCT-efficient scheduling scheme to create a list $S_t$ by sorting all the stages by their job's estimated remaining duration (lines 1-4). The estimation is obtained by calculating the mean of job's duration distribution using the CPDs and the calibration mechanism described in Eq. \eqref{calibration}. 

Next, we adopt the uncertainty reduction strategy to create another list $S_u$ of stages based on their quantified uncertainty reduction (lines 5-10). Due to the uncertainty of compound LLM applications, the duration distribution of a job may overlap with others, potentially leading to errors in sorting jobs by their estimated duration. If the interval of each job's duration distribution doesn't overlap with others, we can ensure 100\% precision in sorting. To this end, we first group all the jobs into multiple non-overlapping sets according to the interval of their duration distribution (line 5). For each non-overlapping set ordered by its lower bound, we sort the stages of all jobs within by their uncertainty reduction and then sequentially add the sorted stages to $S_u$ (lines 6-10). 

We now have two distinct sorted lists of stages: one ordered by estimated job duration and the other by uncertainty reduction. These lists reflect the dual goals of our two scheduling strategies: minimizing average JCT and reducing job uncertainty. To address potential scheduling conflicts, we utilize the $\epsilon$-greedy algorithm \cite{bandit} to combine the two strategies together in an exploration-exploitation manner (lines 11-20). Specifically, at each iteration, we schedule the stage in $S_u$ with a probability of $\epsilon$, and the stage in $S_t$ with a probability of $1-\epsilon$ (lines 13-18). This design balances exploiting current information to reduce average JCT and exploring new information to decrease uncertainty. However, if the stage selected for exploration contains too many tasks, it may significantly delay the execution of small jobs, harming the average JCT. Therefore, we sample a fraction $r$ for scheduling rather than all tasks (line 15). This allows us to estimate the entire duration of the stage by running only a small fraction of tasks. Finally, we attach the rest of the tasks to the end of $T_r$ and $T_l$ and return them as the final scheduling preference lists (lines 21-22). For multiple LLM executors, we use a simple load-balancing method that assigns the next scheduled LLM task to the instance with the fewest running tasks, which scales well with the type and number of LLM compound applications.

{\textbf{Complexity Analysis:}
Assume we have $N$ stages in the system when executing Algorithm \ref{schedulingalgorithm}. Sorting jobs according to their estimated remaining durations (line 1) would take $O(N \log_2 N)$. Gathering all schedulable stages (lines 2-4) requires $O(N)$. Finding the non-overlapping sets of jobs takes $O(N)$ (line 5). Sorting the stages according to their uncertainty reduction (lines 6-10) takes $O(N \log_2 N)$. Performing task sampling for all stages (lines 11-20) takes $O(N)$. As a consequence, the time complexity of Algorithm \ref{schedulingalgorithm} is $O(N \log_2 N)$, which is scalable to the vast type and number of LLM compound applications.} 

{Note that the analysis above omits the overhead of Bayesian inference for calculating the remaining job duration (line 1) and uncertainty reduction of each stage (line 8). This omission is justified for two reasons: First, the Bayesian inference can be reduced to $O(1)$ by keeping a lookup table for all the possible combinations of discrete values of different stages. Second, to optimize inference costs and improve efficiency, existing LLM compound applications are typically designed with fewer than 10 \textit{LLM stages} \cite{taskbench} \cite{autogen} \cite{llmcompiler}, which constitute the majority of nodes in a BN, and not all stages are included in it. It makes Bayesian inference on such small networks effectively constant time. The scheduling latency recorded in our testbed experiment (Sec. \ref{section testbed results}) confirms this low overhead.}

\section{Evaluation}
\label{experiment}

\textbf{Implementation:} 
We implement LLMSched using Python 3.9 with 3k lines of code. We record the runtime duration of different compound LLM applications on the corresponding dataset, and use them to train Bayesian networks with PyAgrum \cite{pyagrum} for all applications except for task automation since it only has one \textit{LLM stage} and one \textit{dynamic stage}. The duration distribution of each stage is discretized into up to 6 intervals based on its frequency. For the LLM, we use the Llama2-7B model \cite{llama} with vLLM \cite{vllm} as the serving engine.
\\
\textbf{Workload generation:} We choose 6 representative compound LLM applications from three categories mentioned in Sec. \ref{background}: (1) Predefined applications: sequence sorting \cite{got} with synthetic dataset and document merging \cite{got} with the dataset from the paper. (2) Chain-like applications: code generation \cite{reflexion} with the MBPP \cite{mbpp} dataset and web search \cite{react} with the HotpotQA \cite{hotpotqa} dataset. (3) Planning applications: task automation \cite{taskbench} with the TaskBench \cite{taskbench} dataset and LLMCompiler \cite{llmcompiler} with the HotpotQA \cite{hotpotqa} dataset.

Using the six compound LLM applications, we create four types of workloads as follows: (1) \textbf{Mixed workload} with the numbers of jobs uniformly distributed across the six applications. (2) \textbf{Predefined workload} with 50\% sequence sorting and 50\% document merging. (3) \textbf{Chain-like workload} with 50\% code generation and 50\% web search. (4) \textbf{Planning workload} with 50\% task automation and 50\% LLMCompiler.

For each type of workload, we randomly sample queries from the corresponding dataset and make the arrival of jobs follow a Poisson distribution with an arrival rate $\lambda$ \cite{decima}.\\
\textbf{Testbed:} We conduct the real experiment on a server equipped with an NVIDIA H800-100G GPU card, an AMD EPYC 9654 Processor, and 120 GB memory. 
\\
\textbf{Simulator:} We also build a simulator based on large-scale evaluation. {We empirically found that batch size primarily influences decoding latency, while other factors, such as the number of batched tokens, have little impact on it. Thus, we profile the relationship between the decoding latency and the batch sizes. Based on the profile, we dynamically adjust the remaining duration of each running LLM task whenever the number of concurrent running requests changes.}
\\
\textbf{Baselines:} We compare the performance of LLMSched against the following six baselines:
\begin{itemize}
    \item First Come First Serve (FCFS): FCFS schedules the jobs in arrival order and is Spark’s default scheme.
    \item Fair Scheduling (Fair): Fair allocates the same number of resources to each running job and is supported by Spark.
    \item Shortest Job First (SJF): SJF prioritizes the job with the shortest estimated duration.
    \item Argus \cite{argus}: Argus ranks the stages based on their depth, number of children, and number of tasks.
    \item Decima \cite{decima}: Decima leverages reinforcement learning (RL) to learn scheduling policy for DAG jobs. {The RL agent is trained on four types of workload mentioned above for fair comparison.}
    \item Carbyne \cite{carbyne}: Carbyne altruistically schedules the job by collecting and redistributing leftover resources. 
\end{itemize}
For prior information required by the baselines, we calculate the average duration and resource requirements for each application on its dataset and extract the DAG structure from our proposed DAG-based model.
\\
\textbf{Parameter setting:} We create a default parameter setting with 300 jobs and a job arrival rate of 0.9 as used in \cite{flexllm}, and manually configure the resources of LLM executors and regular executors for four types of workloads to achieve a moderate cluster load of 85\% in average as in \cite{decima}.
\subsection{Simulation Results}
\begin{figure*}[h!]
    \centering
    \includegraphics[width=1\textwidth]{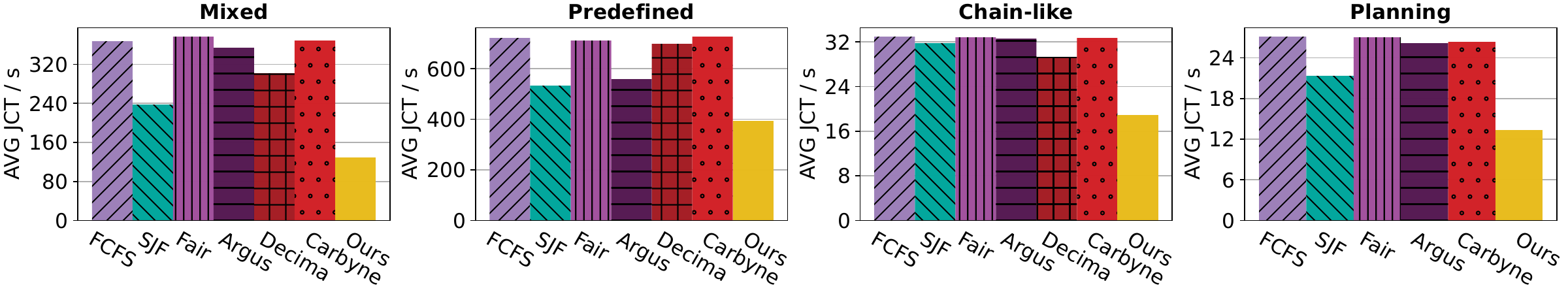}
    \caption{Testbed results of different approaches for four types of workloads ($\lambda=0.9$, 300 jobs). LLMSched is denoted as \textit{Ours}. Decima is omitted in the planning workload because its average JCT is too high (more than 100s).}
    \label{testbed result}
\end{figure*}

 We present the simulation results for different numbers of jobs in Fig. \ref{simulation result}. The results show that LLMSched outperforms all other baselines in terms of average JCT. LLMSched can reduce the average JCT by 36$\sim$79\%, 14$\sim$46\%, 36$\sim$67\%, and 24$\sim$52\% for four types of workloads, respectively. In addition, the advantage of LLMSched becomes more pronounced as the number of jobs increases. For example, compared to Decima, LLMSched reduces the average JCT by 38\%, 65\%, 73\%, and 75\% for the mixed workload with 100, 200, 300, and 400 jobs, respectively, which demonstrates the scalability of LLMSched. 
 
 {We now analyze the performance of different methods under four types of workload in detail. Job-agnostic methods, such as FCFS, and fairness-based methods, like Fair and Carbyne, perform poorly across all four workload types because they fail to optimize the average JCT, which is the most critical factor for compound LLM applications. For Mixed workloads, due to the significant variation in job durations across different compound LLM applications, SJF outperforms other baselines by prioritizing jobs with shorter durations. However, SJF still lags behind LLMSched because it does not account for the uncertainty in LLM compound applications, which can lead to errors in job prioritization. For Predefined workloads, , Argus can efficiently schedule jobs due to its topology awareness, as these workloads have a fixed DAG structure. However, as the job of the same application shares the same topology in Predefined workloads, Argus actually performs application-level scheduling, which is suboptimal in terms of average JCT due to duration uncertainty. In contrast, LLMSched can leverage BN to update the remaining duration estimation and determine job priority within the same application, thus outperforming Argus.}
 
 {For Chain-like workloads, which involve both duration and structural uncertainty, LLMSched outperforms other baselines by a significant margin due to its advantage in handling intrinsic uncertainty in compound LLM applications. For Planning workloads, where jobs exhibit  high stage parallelism (e.g., multiple parallel tool invocations in LLMCompiler \cite{llmcompiler}) but low task parallelism (e.g., each stage contains only one task), Decima performs worst because it schedules tasks of only one stage at a time, leaving resource under-utilized. LLMSched still performs the best, as its exploration-exploitation strategy first reveals the \textit{dynamic stage} of jobs to mitigate uncertainty, and then uses the obtained information for more efficient scheduling.}

\begin{table}[]
\centering
\caption{Average scheduling overhead of each method (ms).}
\begin{tabular}{>{\centering\arraybackslash}m{1.2cm} >{\centering\arraybackslash}m{1.2cm} >{\centering\arraybackslash}m{1.2cm} >{\centering\arraybackslash}m{1.5cm}>{\centering\arraybackslash}m{1.2cm}}
    \toprule
    \multicolumn{1}{c}{} & \textbf{Mixed} & \textbf{Predefined} & \textbf{Chain-like} & \textbf{Planning} \\ \toprule
    \textbf{FCFS} & 0.25 & 0.47& 0.06& 0.1\\ 
    \textbf{SJF} & 0.41& 0.66& 0.17& 0.18 \\ 
    \textbf{FAIR} & 0.48& 0.7& 0.08& 0.12 \\ 
    \textbf{Argus} & 0.32& 0.53& 0.07& 0.1 \\ 
    \textbf{Decima} & 17.7& 28.79& 12.68& 10.17 \\
    \textbf{Carbyne} & 4.39& 8.23& 0.6& 0.62 \\ 
    \textbf{LLMSched} & 0.96& 2.32& 0.7& 0.16 \\ \bottomrule
\end{tabular}
\label{schedulingoverhead}
\end{table}
\subsection{Testbed Results}
\label{section testbed results}
Fig. \ref{testbed result} shows the average JCT of each method for four types of workloads on our testbed. Compared to the other six baselines, LLMSched reduces the average JCT by 45$\sim$66\%, 26$\sim$46\%, 35$\sim$45\%, and 38$\sim$51\% for four types of workloads, respectively. The comparison results on our testbed are nearly consistent with those on the simulation environments, which validates the accuracy of our simulator.

We also present the average scheduling overhead of each method in Table \ref{schedulingoverhead} during our testbed experiment. The results are calculated by dividing the total overhead by the number of invocations for each method. Note that LLMSched's overhead also includes BN inference and entropy calculation. The average scheduling overhead of our method is slightly higher than simple heuristics like FCFS, SJF, and Fair, but much lower than two complex methods, Decima and Carbyne. In addition, the average overhead of LLMSched is under 3 ms for all types of workloads, indicating that LLMSched can perform efficient real-time scheduling without impacting the average JCT.

\begin{figure*}
    \centering
    \subfloat[]{
    \includegraphics[width=0.31\textwidth]{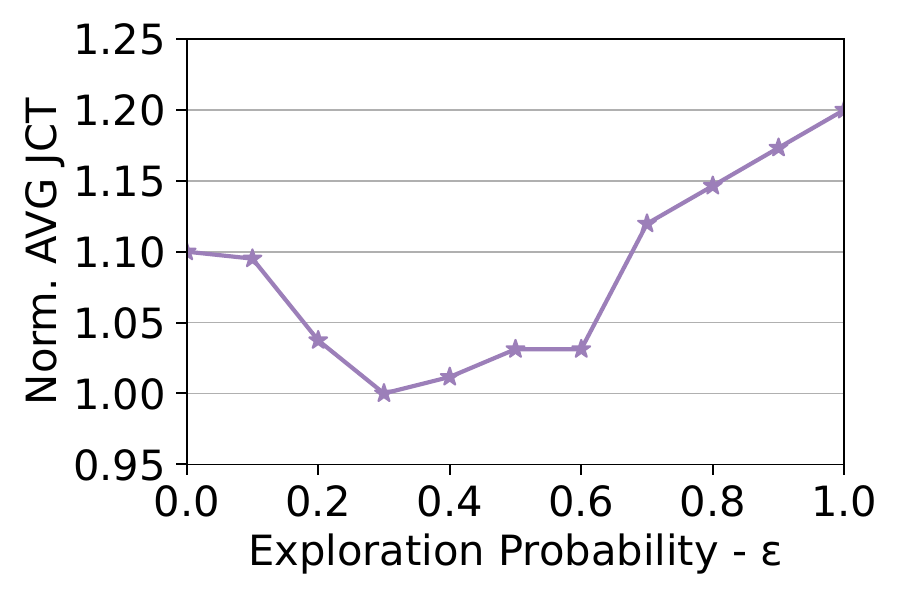}
        \label{epsilon}
    }
    \subfloat[]{
        \includegraphics[width=0.31\textwidth]{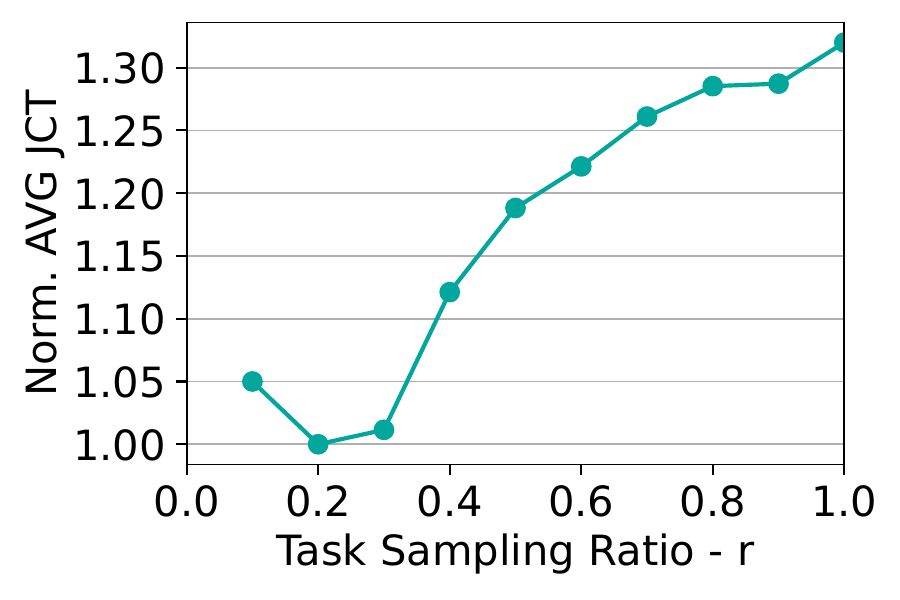}
        \label{tasksamplingratio}
    }
    \subfloat[]{
        \includegraphics[width=0.31\textwidth]{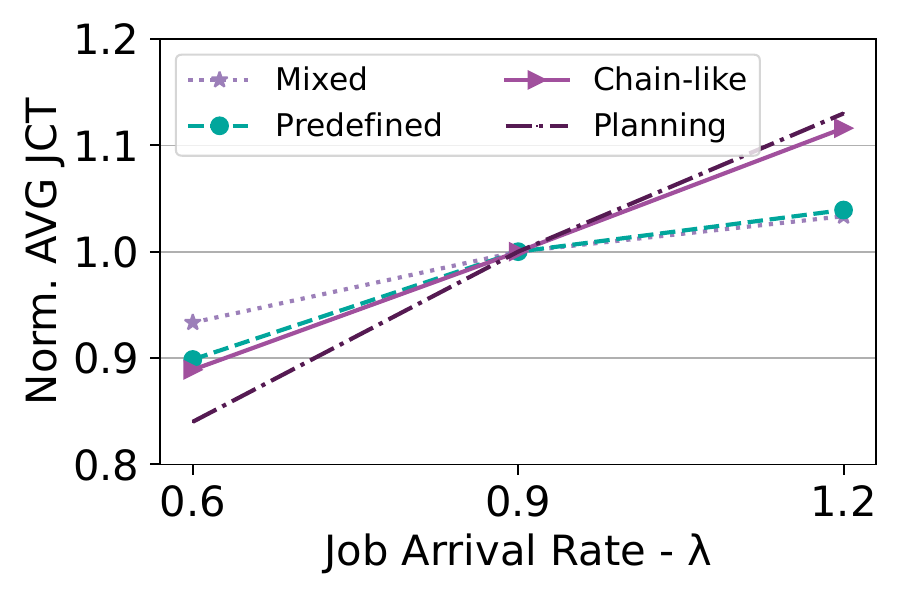}
        \label{lambda}
    }
    \caption{Sensitivity analysis of LLMSched under different exploration probability, task sampling ratio, and job arrival rate.}
    \label{sensitivity}
\end{figure*}

\begin{figure}
    \centering
    \includegraphics[width=0.48\textwidth]{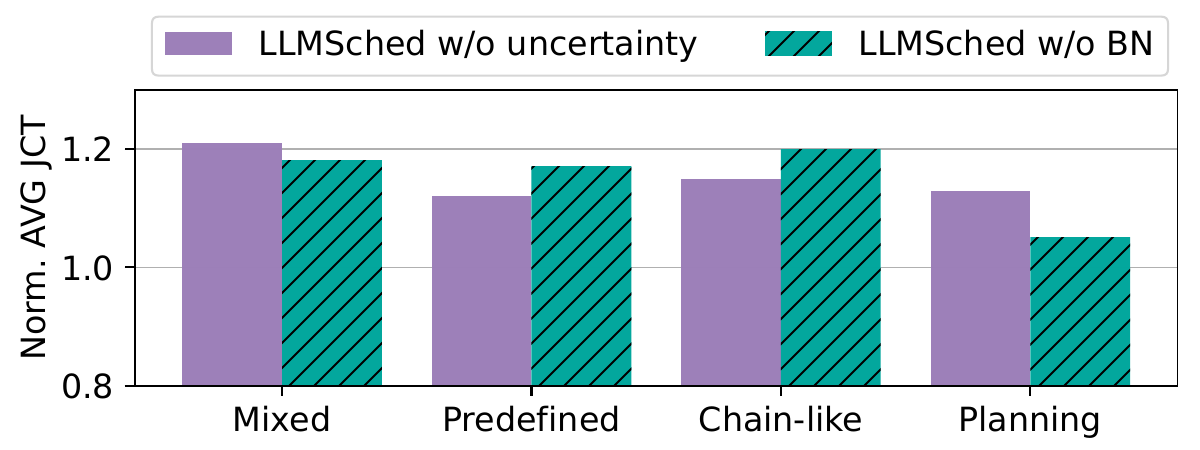}
    \caption{Ablation study of LLMSched on four types of workloads.}
    \label{ablation}
\end{figure}

\subsection{Ablation Study}

Our LLMSched framework comprises two critical components, one is the Bayesian network for updating the estimation of the descendant stages' distributions, and the other is the uncertainty-aware scheduling strategy. To analyze the effectiveness of both components, we conduct an ablation study. We create two additional methods, \textit{LLMSched w/o BN} and \textit{LLMSched w/o uncertainty}. The first one follows the same scheduling schemes presented in Algorithm \ref{schedulingalgorithm} but uses the average historical job duration for estimation. The second one uses the Bayesian Network to update the posterior distribution of job duration but only performs the SRTF strategy.

Fig. \ref{ablation} shows the results of the ablation study on four types of workloads. We normalize the average JCT of the two methods to that of LLMSched for comparison purposes. {For \textit{LLMSched w/o BN}, the average JCT is 18\%, 17\%, 20\%, and 5\% higher than LLMSched on four types of workloads. This indicates that BN plays an important role as it significantly enhances the accuracy of job duration estimation. With BN, the job duration estimation for unscheduled stages can be dynamically updated and predicted more accurately by leveraging inter-stage correlations. For \textit{LLMSched w/o uncertainty}, the average JCT is 21\%, 12\%, 15\%, and 13\% higher than LLMSched on four types of workloads, respectively. This suggests that the uncertainty-aware strategy is crucial in effectively guiding the exploration process. Its importance becomes particularly prominent when handling the Mixed workload, where uncertainty reduction varies significantly across stages. For this workload, the performance of \textit{LLMSched w/o BN} surpasses that of \textit{LLMSched w/o uncertainty}.}

\subsection{Sensitivity Analysis}
In this section, we analyze the performance of LLMSched under different exploration probability $\epsilon$, task sampling ratio $r$, and job arrival rate $\lambda$. Fig. \ref{epsilon} shows the relationship between the exploration probability and the normalized average JCT. As the probability increases, the average JCT initially decreases and then gradually increases. It indicates a trade-off between exploration and exploitation, suggesting that a balance should be struck between them. 

Fig. \ref{tasksamplingratio} presents the relationship between task sampling ratio and normalized average JCT. When the ratio is too small, the error of estimation of job duration can be large and thus it increases the average JCT. When the ratio becomes larger, more resources are allocated for task sampling and the small jobs will be delayed, resulting in higher average JCT. Therefore, a suitably small task sampling ratio is preferred.

Fig. \ref{lambda} shows the normalized JCT corresponding to job arrival rate with a value of 0.6, 0.9 and 1.2 for four types of workload, which simulate the lightly-loaded, moderately loaded, and heavily-loaded cluster environment. Generally, the average JCT increases as the job arrives more frequently. The average JCT of LLMSched changes more significantly for chain-like and planning workloads, and more smoothly for mixed and predefined workloads.

\section{Related Work}
\label{relatedwork}

\subsection{DAG Job Scheduling}
Scheduling data processing jobs with the directed acyclic graph (DAG) structure to reduce average job completion time (JCT) or makespan has been widely studied. A majority of existing works \cite{graphene} \cite{decima} \cite{branch} schedule jobs by leveraging sufficient prior knowledge (i.e., DAG structure and job duration) for recurring jobs. Graphene \cite{graphene} prioritizes troublesome jobs that are long-running or hard to pack. Decima \cite{decima} applies a graph neural network to capture DAG topology and reinforcement learning to learn the scheduling policy. Branch scheduling \cite{branch} organizes DAG stages into branches to reduce data transfer overhead within each job. These methods rely on accurate information about job duration and DAG structure, which is often unavailable in compound LLM applications. Our framework efficiently handles the uncertainty in compound LLM applications and successfully improves their scheduling performance.

\subsection{LLM Serving}
To accelerate the inference speed and increase the throughput of LLM, several works have been proposed \cite{orca} \cite{fastserve} \cite{flashattention} \cite{vllm}. Orca \cite{orca} schedules LLM requests at the iteration-level rather than the request-level and performs selective batching. FastServe \cite{fastserve} uses preemptive scheduling to minimize response time with a novel skip-join Multi-Level Feedback Queue scheduler. FlashAttention \cite{flashattention} uses tiling to reduce the number of memory reads/writes to boost LLM inference speed. vLLM \cite{vllm} adopts a novel PagedAttention to manage the memory usage of the KV cache in LLM and improve the throughput. While these works focus only on single-request serving, we present the first uncertainty-aware scheduling framework tailored for emerging compound LLM applications. 

\section{Conclusion}
\label{conclusion}
In this paper, we propose LLMSched, an uncertainty-aware scheduling framework for compound LLM applications. We conduct data analysis on current representative compound LLM applications, revealing their duration and structural uncertainty and the challenges in scheduling. Then, we design a novel DAG-based model and Bayesian network-based profiler to capture the uncertainty, and incorporate an entropy-based mechanism to quantify it. Next, we propose an uncertainty-aware scheduling scheme to reduce the average JCT. Experimental results from both the simulation environment and testbed demonstrate the superiority of our method over state-of-the-art scheduling algorithms.

\section{Acknowledgment}
This work is supported by the National Key R\&D Program of China (Grant No. 2024YFC3017100).

\bibliography{ref.bib}

\begin{thebibliography}{10}
\providecommand{\url}[1]{#1}
\csname url@samestyle\endcsname
\providecommand{\newblock}{\relax}
\providecommand{\bibinfo}[2]{#2}
\providecommand{\BIBentrySTDinterwordspacing}{\spaceskip=0pt\relax}
\providecommand{\BIBentryALTinterwordstretchfactor}{4}
\providecommand{\BIBentryALTinterwordspacing}{\spaceskip=\fontdimen2\font plus
\BIBentryALTinterwordstretchfactor\fontdimen3\font minus \fontdimen4\font\relax}
\providecommand{\BIBforeignlanguage}[2]{{%
\expandafter\ifx\csname l@#1\endcsname\relax
\typeout{** WARNING: IEEEtran.bst: No hyphenation pattern has been}%
\typeout{** loaded for the language `#1'. Using the pattern for}%
\typeout{** the default language instead.}%
\else
\language=\csname l@#1\endcsname
\fi
#2}}
\providecommand{\BIBdecl}{\relax}
\BIBdecl

\bibitem{gpt4}
J.~Achiam, S.~Adler, S.~Agarwal, L.~Ahmad, I.~Akkaya, F.~L. Aleman, D.~Almeida, J.~Altenschmidt, S.~Altman, S.~Anadkat \emph{et~al.}, ``Gpt-4 technical report,'' \emph{arXiv preprint arXiv:2303.08774}, Mar. 2023.

\bibitem{gemini}
G.~Team, R.~Anil, S.~Borgeaud, Y.~Wu, J.-B. Alayrac, J.~Yu, R.~Soricut, J.~Schalkwyk, A.~M. Dai, A.~Hauth \emph{et~al.}, ``Gemini: a family of highly capable multimodal models,'' \emph{arXiv preprint arXiv:2312.11805}, Dec. 2023.

\bibitem{llama}
H.~Touvron, T.~Lavril, G.~Izacard, X.~Martinet, M.-A. Lachaux, T.~Lacroix, B.~Rozi{\`e}re, N.~Goyal, E.~Hambro, F.~Azhar \emph{et~al.}, ``Llama: Open and efficient foundation language models,'' \emph{arXiv preprint arXiv:2302.13971}, Feb. 2023.

\bibitem{vicuna}
\BIBentryALTinterwordspacing
(2023) Vicuna: An open-source chatbot impressing gpt-4 with 90\% chatgpt quality. [Online]. Available: \url{https://lmsys.org/blog/2023-03-30-vicuna/}
\BIBentrySTDinterwordspacing

\bibitem{qwen}
J.~Bai, S.~Bai, Y.~Chu, Z.~Cui, K.~Dang, X.~Deng, Y.~Fan, W.~Ge, Y.~Han, F.~Huang \emph{et~al.}, ``Qwen technical report,'' \emph{arXiv preprint arXiv:2309.16609}, Sep. 2023.

\bibitem{gptstatistics}
\BIBentryALTinterwordspacing
(2024) 107 up-to-date chatgpt statistics \& user numbers [april 2024]. [Online]. Available: \url{https://nerdynav.com/chatgpt-statistics/}
\BIBentrySTDinterwordspacing

\bibitem{llmcompound}
Y.~Zhu, B.~Zhu, C.~Chen, and X.~Fan, ``Towards efficient compound large language model system serving in the wild,'' in \emph{Proc. of the Int. Symp. on Qual. of Serv. (IWQoS)}, Jun. 2024, pp. 1--2.

\bibitem{got}
M.~Besta, N.~Blach, A.~Kubicek, R.~Gerstenberger, M.~Podstawski, L.~Gianinazzi, J.~Gajda, T.~Lehmann, H.~Niewiadomski, P.~Nyczyk \emph{et~al.}, ``Graph of thoughts: Solving elaborate problems with large language models,'' in \emph{Proc. AAAI Conf. on Artif. Intel. (AAAI)}, Mar. 2024, pp. 17\,682--17\,690.

\bibitem{reflexion}
N.~Shinn, F.~Cassano, A.~Gopinath, K.~Narasimhan, and S.~Yao, ``Reflexion: Language agents with verbal reinforcement learning,'' in \emph{Annu. Conf. on Neural Inf. Process. Systs. (NeurIPS)}, Dec. 2024.

\bibitem{taskbench}
Y.~Shen, K.~Song, X.~Tan, W.~Zhang, K.~Ren, S.~Yuan, W.~Lu, D.~Li, and Y.~Zhuang, ``Taskbench: Benchmarking large language models for task automation,'' \emph{arXiv preprint arXiv:2311.18760}, Nov. 2023.

\bibitem{react}
S.~Yao, J.~Zhao, D.~Yu, N.~Du, I.~Shafran, K.~Narasimhan, and Y.~Cao, ``{ReAct}: Synergizing reasoning and acting in language models,'' in \emph{Int. Conf. on Learn. Represent. (ICLR)}, May 2023.

\bibitem{autogen}
Q.~Wu, G.~Bansal, J.~Zhang, Y.~Wu, S.~Zhang, E.~Zhu, B.~Li, L.~Jiang, X.~Zhang, and C.~Wang, ``Autogen: Enabling next-gen llm applications via multi-agent conversation framework,'' \emph{arXiv preprint arXiv:2308.08155}, Aug. 2023.

\bibitem{decima}
H.~Mao, M.~Schwarzkopf, S.~B. Venkatakrishnan, Z.~Meng, and M.~Alizadeh, ``Learning scheduling algorithms for data processing clusters,'' in \emph{Proc. of the ACM Spec. Int. Group on Data Commun. (ACM SIGCOMM)}, Aug. 2019, pp. 270--288.

\bibitem{branch}
Z.~Hu, D.~Li, Y.~Zhang, D.~Guo, and Z.~Li, ``Branch scheduling: Dag-aware scheduling for speeding up data-parallel jobs,'' in \emph{Proceedings of the international symposium on quality of service}, Aug. 2019, pp. 1--10.

\bibitem{orca}
G.-I. Yu, J.~S. Jeong, G.-W. Kim, S.~Kim, and B.-G. Chun, ``Orca: A distributed serving system for transformer-based generative models,'' in \emph{USENIX Symp. Operating Syst. Des. Implementation (OSDI)}, Oct. 2022, pp. 521--538.

\bibitem{fastserve}
B.~Wu, Y.~Zhong, Z.~Zhang, G.~Huang, X.~Liu, and X.~Jin, ``Fast distributed inference serving for large language models,'' \emph{arXiv preprint arXiv:2305.05920}, May 2023.

\bibitem{vllm}
W.~Kwon, Z.~Li, S.~Zhuang, Y.~Sheng, L.~Zheng, C.~H. Yu, J.~Gonzalez, H.~Zhang, and I.~Stoica, ``Efficient memory management for large language model serving with pagedattention,'' in \emph{Proc. 29th Symp. Operating Syst. Princ. (SOSP)}, Oct. 2023, pp. 611--626.

\bibitem{hugginggpt}
Y.~Shen, K.~Song, X.~Tan, D.~Li, W.~Lu, and Y.~Zhuang, ``Hugginggpt: Solving ai tasks with chatgpt and its friends in hugging face,'' in \emph{Annu. Conf. on Neural Inf. Process. Systs. (NeurIPS)}, Dec. 2024.

\bibitem{graphene}
R.~Grandl, S.~Kandula, S.~Rao, A.~Akella, and J.~Kulkarni, ``Graphene: Packing and dependency-aware scheduling for data-parallel clusters,'' in \emph{USENIX Symp. Operating Syst. Des. Implementation (OSDI)}, Nov. 2016, pp. 81--97.

\bibitem{carbyne}
R.~Grandl, M.~Chowdhury, A.~Akella, and G.~Ananthanarayanan, ``Altruistic scheduling in multi-resource clusters,'' in \emph{USENIX Symp. Operating Syst. Des. Implementation (OSDI)}, Nov. 2016, pp. 65--80.

\bibitem{sjf}
K.~Agrawal, J.~Li, K.~Lu, and B.~Moseley, ``Scheduling parallel dag jobs online to minimize average flow time,'' in \emph{Proc. 27th Annu. ACM-SIAM Symp. Discrete Algorithms (ACM-SIAM SDA)}, Oct. 2016, pp. 176--189.

\bibitem{argus}
S.~Wu, H.~Chen, Y.~Wang, and H.~Jin, ``Argus: Efficient job scheduling in rdma-assisted big data processing,'' in \emph{IEEE Int. Parall. and Distrib. Process. Symp. (IPDPS)}, May 2021, pp. 827--836.

\bibitem{langchain}
\BIBentryALTinterwordspacing
(2024) Langchain - build context-aware reasoning applications. [Online]. Available: \url{https://python.langchain.com/}
\BIBentrySTDinterwordspacing

\bibitem{spark}
\BIBentryALTinterwordspacing
(2024) Apache spark. [Online]. Available: \url{https://spark.apache.org/}
\BIBentrySTDinterwordspacing

\bibitem{hadoop}
\BIBentryALTinterwordspacing
(2024) Apache hadoop. [Online]. Available: \url{https://hadoop.apache.org/}
\BIBentrySTDinterwordspacing

\bibitem{codebench}
M.~M. Rahman, A.~Kundu, and E.~Bertino, ``Benchmarking of code generative llms,'' in \emph{2024 IEEE 44th Int. Conf. on Dist. Comp. Syst. (ICDCS)}.\hskip 1em plus 0.5em minus 0.4em\relax IEEE, Jul. 2024, pp. 1448--1449.

\bibitem{llmcompiler}
S.~Kim, S.~Moon, R.~Tabrizi, N.~Lee, M.~Mahoney, K.~Keutzer, and A.~Gholami, ``An llm compiler for parallel function calling,'' \emph{arXiv}, Oct. 2023.

\bibitem{mbpp}
J.~Austin, A.~Odena, M.~Nye, M.~Bosma, H.~Michalewski, D.~Dohan, E.~Jiang, C.~Cai, M.~Terry, Q.~Le \emph{et~al.}, ``Program synthesis with large language models,'' \emph{arXiv preprint arXiv:2108.07732}, Aug. 2021.

\bibitem{bayesiannetwork}
N.~Friedman, D.~Geiger, and M.~Goldszmidt, ``Bayesian network classifiers,'' \emph{Machine learning}, vol.~29, no. 2-3, pp. 131--163, Nov. 1997.

\bibitem{entropy}
C.~E. Shannon, ``A mathematical theory of communication,'' \emph{The Bell system technical journal}, vol.~27, pp. 379--423, Jul. 1948.

\bibitem{dagentropy}
A.~K. Wong and M.~You, ``Entropy and distance of random graphs with application to structural pattern recognition,'' \emph{IEEE Trans. on Pattern Anal. and Mach. Intell. (TPAMI)}, no.~5, pp. 599--609, Sep. 1985.

\bibitem{bandit}
J.-Y. Audibert and S.~Bubeck, ``Best arm identification in multi-armed bandits,'' in \emph{COLT-23th Conference on learning theory-2010}, 2010, pp. 13--p.

\bibitem{pyagrum}
G.~Ducamp, C.~Gonzales, and P.-H. Wuillemin, ``agrum/pyagrum: a toolbox to build models and algorithms for probabilistic graphical models in python,'' in \emph{Int. Conf. on Probabil. Graph. Models (ICPGM)}, Sep. 2020, pp. 609--612.

\bibitem{hotpotqa}
Z.~Yang, P.~Qi, S.~Zhang, Y.~Bengio, W.~W. Cohen, R.~Salakhutdinov, and C.~D. Manning, ``Hotpotqa: {A} dataset for diverse, explainable multi-hop question answering,'' in \emph{Proc. of the Conf. on Empir. Methods in Nat. Lang. Process. (EMNLP)}, Oct. 2018, pp. 2369--2380.

\bibitem{flexllm}
X.~Miao, G.~Oliaro, X.~Cheng, M.~Wu, C.~Unger, and Z.~Jia, ``Flexllm: A system for co-serving large language model inference and parameter-efficient finetuning,'' \emph{arXiv preprint arXiv:2402.18789}, Feb. 2024.

\bibitem{flashattention}
T.~Dao, D.~Fu, S.~Ermon, A.~Rudra, and C.~R{\'e}, ``Flashattention: Fast and memory-efficient exact attention with io-awareness,'' \emph{Annu. Conf. on Neural Inf. Process. Systs. (NeurIPS)}, vol.~35, pp. 16\,344--16\,359, Dec. 2022.

\end{thebibliography}
\bibliographystyle{IEEEtran}

\end{document}